\documentclass[journal]{IEEEtran}

\usepackage{amsmath,epsfig,color}
\usepackage[dvipsnames]{xcolor}
\usepackage{flushend,cite}
\usepackage[hyphens]{url}
\usepackage{upgreek}

\usepackage[utf8]{inputenc}
\usepackage[T1]{fontenc}
\usepackage{amsmath}
\usepackage{graphicx} 
\usepackage{hyperref}
\usepackage{subfig}
\usepackage{float}
\usepackage{caption}
\usepackage[ruled,vlined]{algorithm2e}
\usepackage{xcolor}
\usepackage[thickqspace,cdot,squaren]{SIunits}
\usepackage{balance}

\usepackage{tikz,xcolor,hyperref}
\definecolor{lime}{HTML}{A6CE39}
\DeclareRobustCommand{\orcidicon}{%
	\begin{tikzpicture}
	\draw[lime, fill=lime] (0,0) 
	circle [radius=0.16] 
	node[white] {{\fontfamily{qag}\selectfont \tiny ID}};
	\draw[white, fill=white] (-0.0625,0.095) 
	circle [radius=0.007];
	\end{tikzpicture}
	\hspace{-2mm}
}
\foreach \x in {A, ..., Z}{%
	\expandafter\xdef\csname orcid\x\endcsname{\noexpand\href{https://orcid.org/\csname orcidauthor\x\endcsname}{\noexpand\orcidicon}}
}

\protect

\begin{document}

\title{Automatic Dataset Builder for  Machine Learning Applications to Satellite Imagery} 

\newcommand{\orcidauthorA}{0000-0001-6294-0581}
\newcommand{\orcidauthorB}{0000-0002-9252-907X}

\author{Alessandro Sebastianelli \orcidB{}, Maria Pia Del Rosso, Silvia Liberata Ullo \orcidA{}
\thanks{Alessandro Sebastianelli, Maria Pia Del Rosso, Silvia Liberata Ullo, University of Sannio,   Benevento, Italy (e-mail: {\it sebastianelli@unisannio.it}, {\it mari\nobreak apia.delrosso@gmail.com}, {\it ullo@unisannio.it}}
}

\maketitle

\begin{abstract}
Nowadays the use of Machine Learning (ML) algorithms is spreading in the field of Remote Sensing, with applications ranging from detection and classification of land use and monitoring to the prediction of many natural or anthropic phenomena of interest. One main limit of their employment is related to the need for a huge amount of data  for training the neural network, chosen for the specific application, and the resulting computational weight and time required to collect the necessary data. In this letter the architecture of an innovative tool, enabling researchers to create in an automatic way  suitable datasets for AI (Artificial Intelligence)  applications in  the EO (Earth Observation) context, is presented. Two versions of the architecture have been implemented and made available on Git-Hub, with a specific Graphical User Interface (GUI) for non-expert users.
\end{abstract}


\IEEEpeerreviewmaketitle

\section{Introduction}
In recent years, in the field of Remote Sensing, a large number of applications have benefit by the introduction of Machine Learning and Deep Learning techniques in their data processing workflow.
However, when working with Machine Learning,  the availability of a sufficiently large and statistically representative dataset becomes crucial\cite{5306233, 8862913, data_required, data_impact}. 

\begin{figure}[!ht]
    \centering
    \includegraphics[scale=0.3]{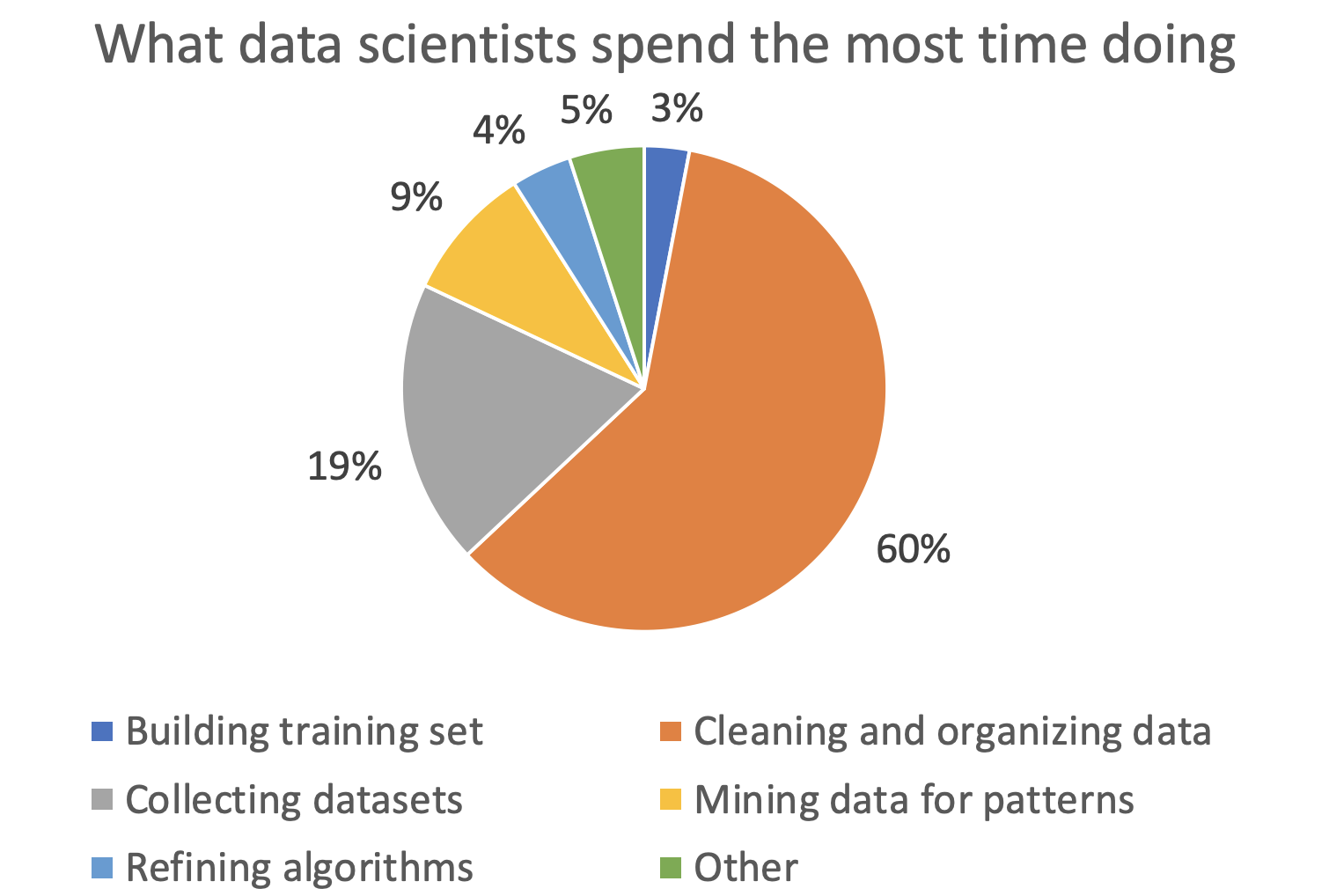}
    \caption{Time consuming activities of data science  \cite{cleaning_big_data}.}
    \label{fig:time1}
\end{figure}

\begin{figure}[!ht]
    \centering
    \includegraphics[scale=0.3]{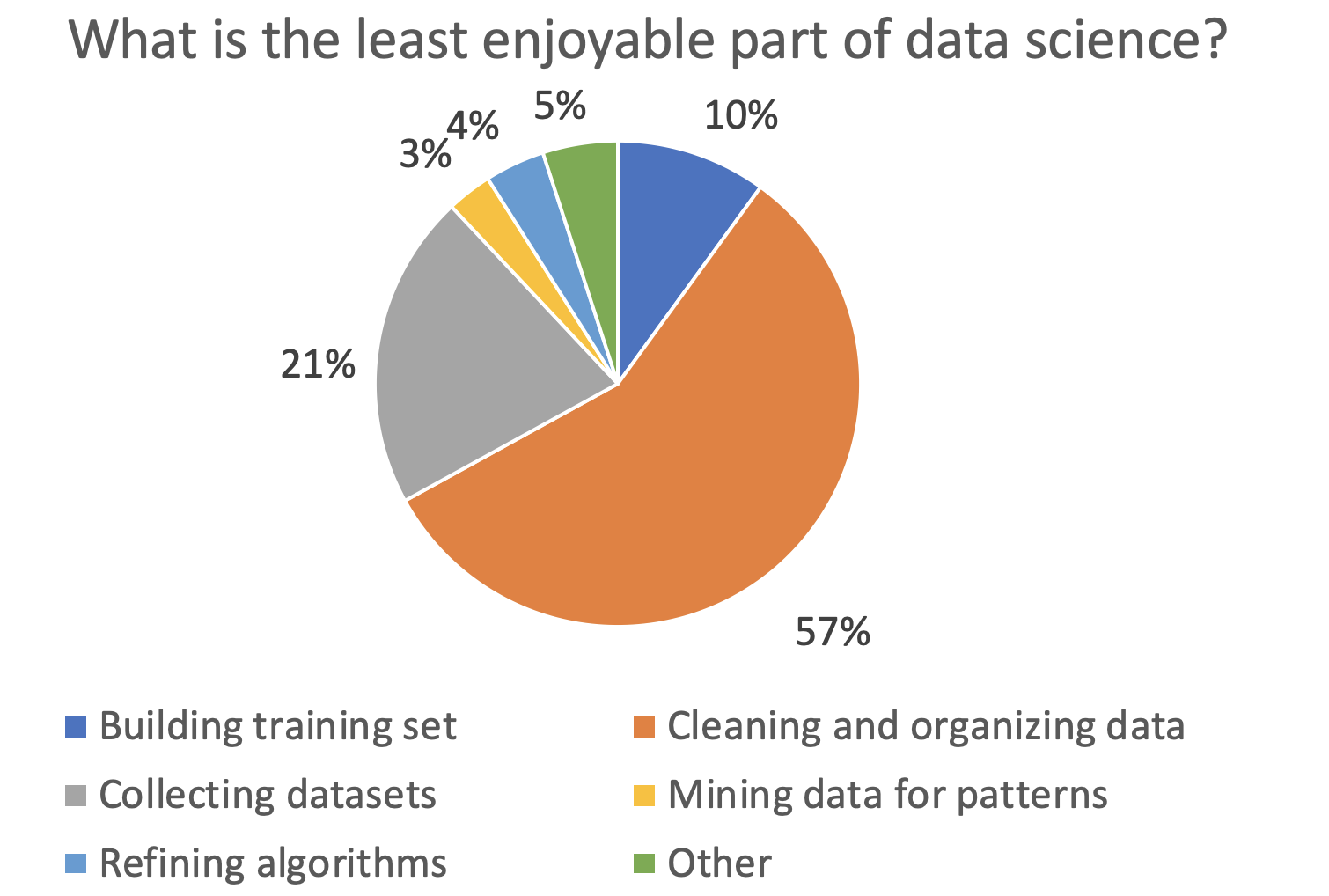}
    \caption{Least enjoyable activities of data science \cite{cleaning_big_data}.}
    \label{fig:time2}
\end{figure}

A huge amount of data  is necessary for training the neural network, chosen for the specific application, and this issue brings out all the problems related to Big Data, and their handling.
The process of creating a dataset is a very slow and laborious operation, which 
forces the researcher to waste more of his time before focusing on the most interesting part of his work. As discussed in \cite{cleaning_big_data}, and shown in the Figure \ref{fig:time1}, collecting, building, organizing and cleaning the data  is a heavy time-consuming operation, which results in researchers' frustration, since it is felt as the least enjoyable part of data processing, as represented in the Figure \ref{fig:time2}. \\
The architecture of an automatic dataset builder is proposed in this letter, with a detailed description of its several blocks, in order to make available an innovative tool enabling researchers to create suitable datasets for  Artificial Intelligence (AI)  applications in  the Earth Observation (EO) context. The annoying and repetitive operations are done in this way by the software in an automatic way, and the researcher can save the time for other activities. 

From the analysis of the state of the art, the availability of such architectures is very limited. A couple of them are present in the literature (\cite{schmitt2018sen1},  \cite{ranghetti2020sen2r}), but the  model  proposed in this paper has   many advantages   with respect to them: the fully automatic chain processing, the possibility to to download and organize time series of data from multiple sources, the presence of a GUI for non-expert users, the possibility, especially for expert users, to add  their pre-processing techniques to the processing chain,   the availability on Git-Hub (open access).

The paper is organized as follow, after the Introduction, the architecture of the dataset builder  is described, with its main components, in the Section II. In the Section III the Graphical User Interface (GUI) designed for the non-expert users is presented. Finally, conclusions are given in the last Section. 

\begin{figure*}[!ht]
    \centering
    \includegraphics[scale=0.455]{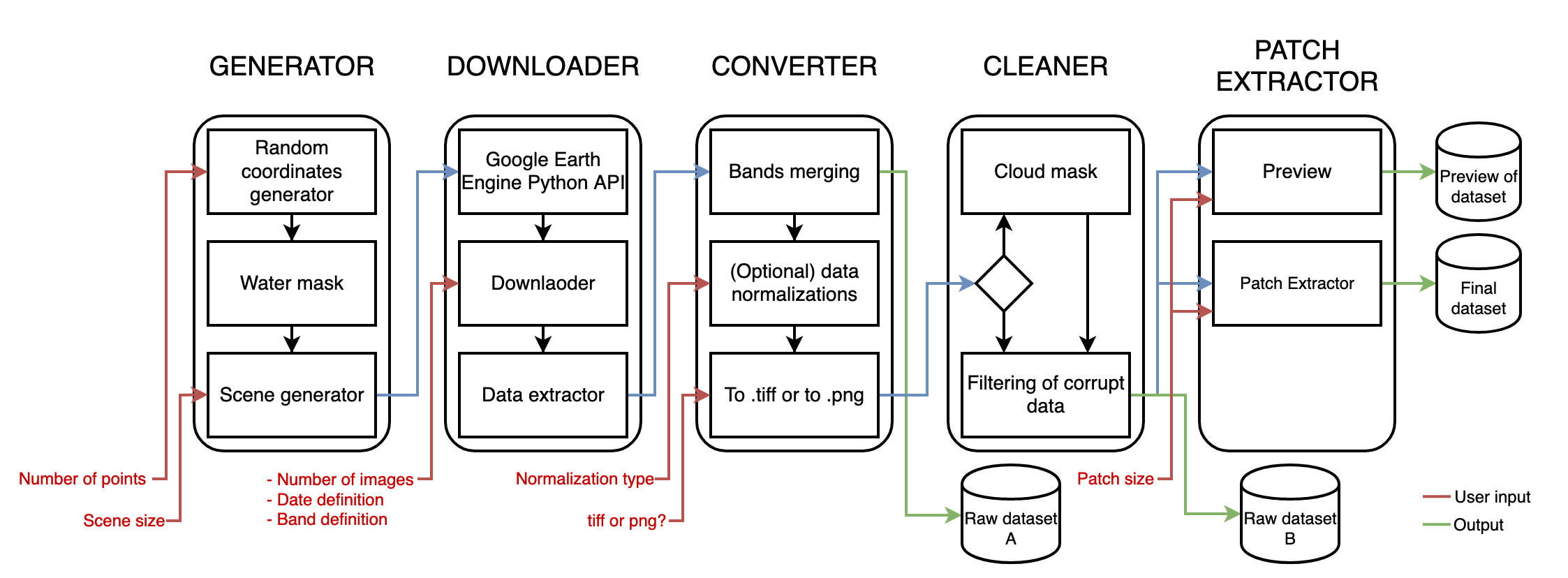}
    \caption{Functional Diagram for the Satellite Dataset Creation Tool}
    \label{fig:diagram_downloader_tool}
\end{figure*}

\section{Description}
As before specified, the proposed architecture describes the tool  designed to build suitable datasets for Machine Learning   applications in a simple and automatic way. A set of Python scripts allows the user to automatically download data from the Google Earth Engine catalog \cite{google_earth_engine}. The Functional Diagram for the Satellite Dataset Creation Tool is shown in the Figure \ref{fig:diagram_downloader_tool}, where the several functional blocks are included, each of them dealing with a particular task. The parameters to be set are few (the red labels in the Figure \ref{fig:diagram_downloader_tool}), and among them: the coordinates of the area of interest, its size, the dates, the data bands, the number of images. In the next subsections the functioning of the various functional blocks will be explained.

\subsection{Generator}\label{subs:generator}
The generator produces points, with longitude and latitude, distributed over the Earth surface. In the case of both Sentinel-1 and Sentinel-2 (the two satellites for which some examples of dataset creation will later on be provided), since no data acquisition is conducted over the seas and oceans, it has been necessary to introduce a watermasking function inside the generation process \cite{global_land_mask}: 

\subsubsection{watermasking} 
\begin{figure}[!ht]
    \centering
    \includegraphics[scale=0.22]{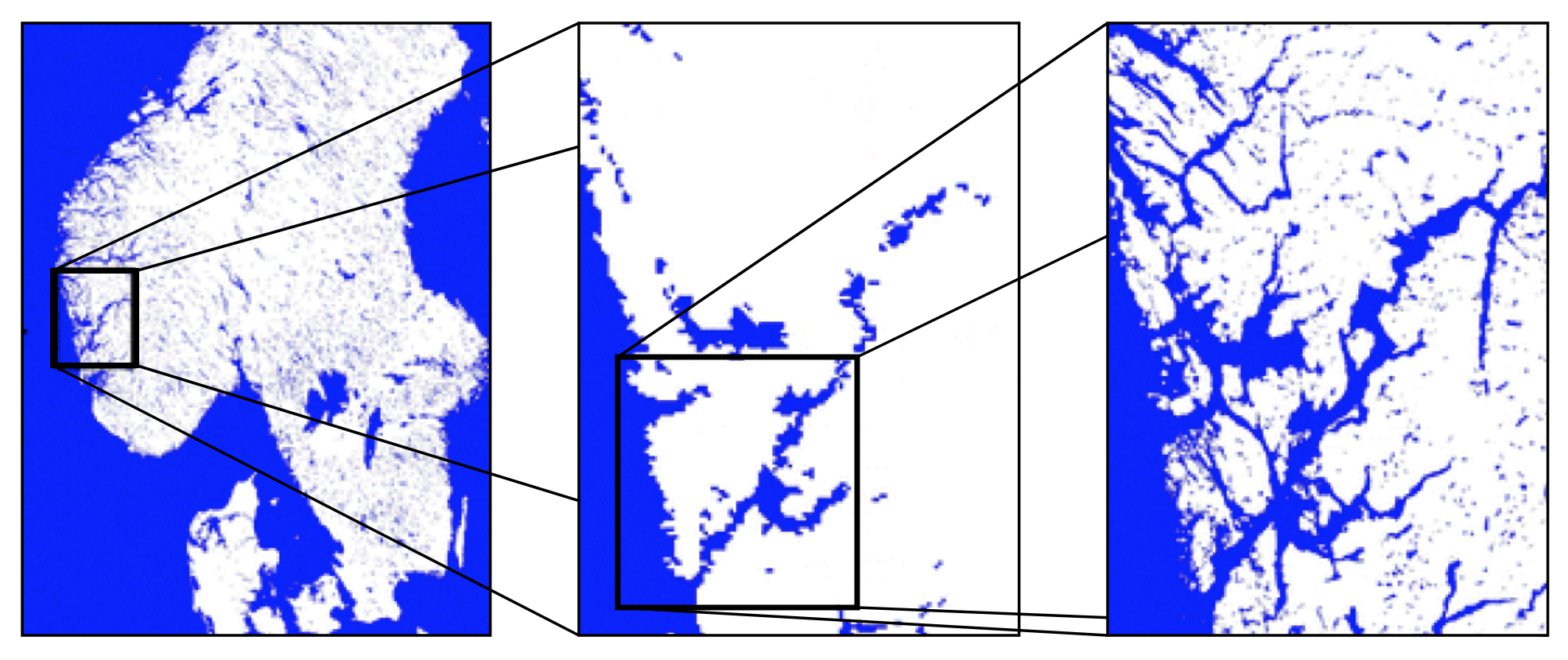}
    \caption{Example of Water Mask}
    \label{fig:water_mask}
\end{figure}

a mask is used that allows identifying and delimiting water rich areas of the Earth with a certain resolution; two classes are identified, water (value 1) and not water or land (value 0), with the blue color  associated to the water, and the white color to the rest,  as shown in the Fig. \ref{fig:water_mask}.

\begin{figure}[!ht]
    \centering
    \includegraphics[scale=0.25]{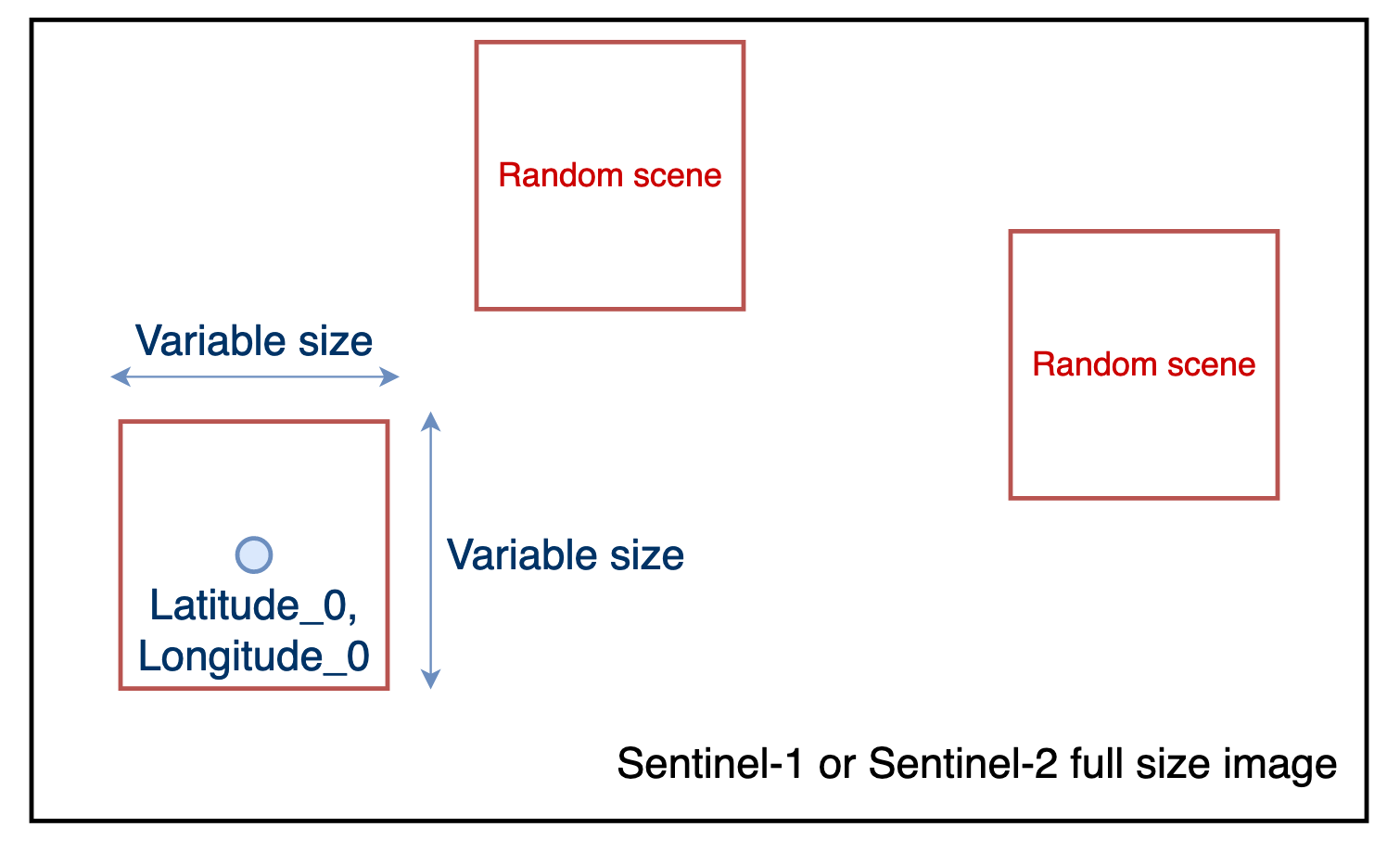}
    \caption{Example of the generator output: a square-shaped list of geo-referenced points.}
    \label{fig:generator_example}
\end{figure}

To generate the points, two random variables with uniform distribution have been used, one for the latitude values and the other for the longitude values. The latitudes are restricted to the range  $[-56, 84]$, since Sentinel-2 does not acquire data beyond those values, while for the longitude the range is restricted to $[-180, 180]$ \cite{sen2_acq_plan, sen1_acq_plan}. At each iteration a point is generated, and through the water mask it is verified if the point falls on the earth’s surface. Only in this case, it is saved in a csv file dedicated to the generator, while it is discarded in negative case. The user can define the number of points to be generated and the size of the scene. The generator output is a square-shaped list of geo-referenced points, see Figure \ref{fig:generator_example}.

\subsection{Downloader}\label{subs:downloader}
The downloader takes care of downloading the images using the coordinates previously generated and the time interval specified by the user. By default the script will try to download a one-year time series, with a monthly interval. For each month three Sentinel-1 images and three Sentinel-2 images are downloaded, this number was chosen to guarantee at least one image for each satellite that is in optimal conditions of light, cloud coverage, etc. The software is also designed to organize data in a hierarchical folder structure, for example for a region the structure is as follows (with the Sentinel-1 and Sentinel-2 data folders both contained in a master folder):

\begin{itemize}
    \item Sentinel-1 (or Sentinel-2) folder:
    \begin{itemize}
        \item Scene 1 folder:
        \begin{itemize}
            \item January folder
            \begin{itemize}
                \item image 1
                \item image 2
                \item image 3
            \end{itemize}
            \item February folder
            \item \dots
        \end{itemize}
        \item Scene 2 folder
        \item \dots
    \end{itemize}
\end{itemize}

The user can define the number of images, the date and the bands for each satellite. The default Sentinel-2 bands are B4, B3, B2 and QA60 (R, G, B and cloud mask) and for Sentinel-1 the default value is VV. The downloader deals exclusively with downloading the raw data, and this is the reason why the converter block becomes necessary after the downloader.

\subsection{Converter}\label{subs:converter}
\begin{figure}[!ht]
    \centering
    \includegraphics[scale=.24]{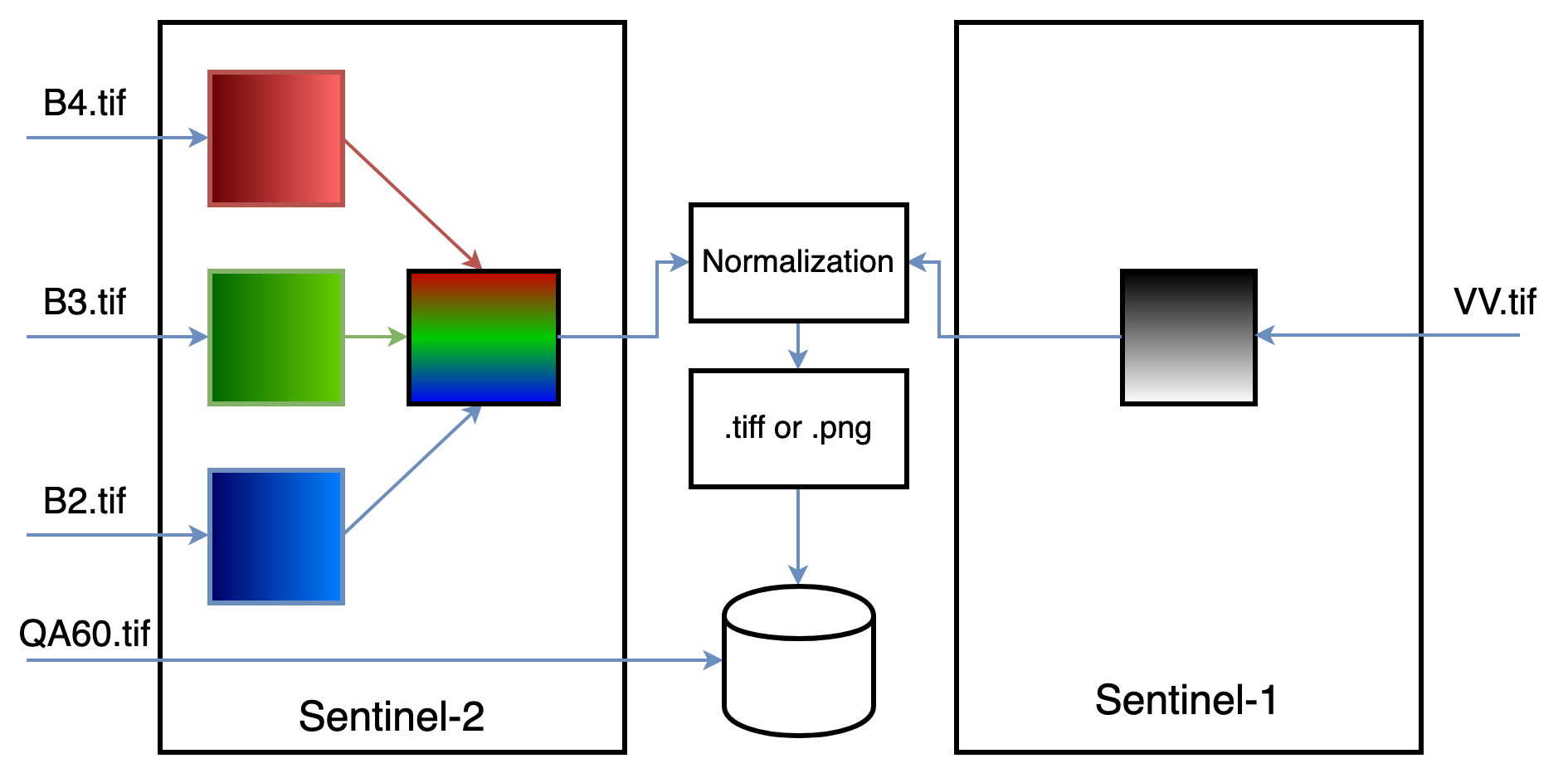}
    \caption{Converter functional scheme}
    \label{fig:converter}
\end{figure}

The converter mainly deals with taking raw data and applying some preprocessing techniques to produce as output easily treatable data, see Figure \ref{fig:converter}.

\begin{figure}[!ht]
    \centering
    \includegraphics[scale=.08]{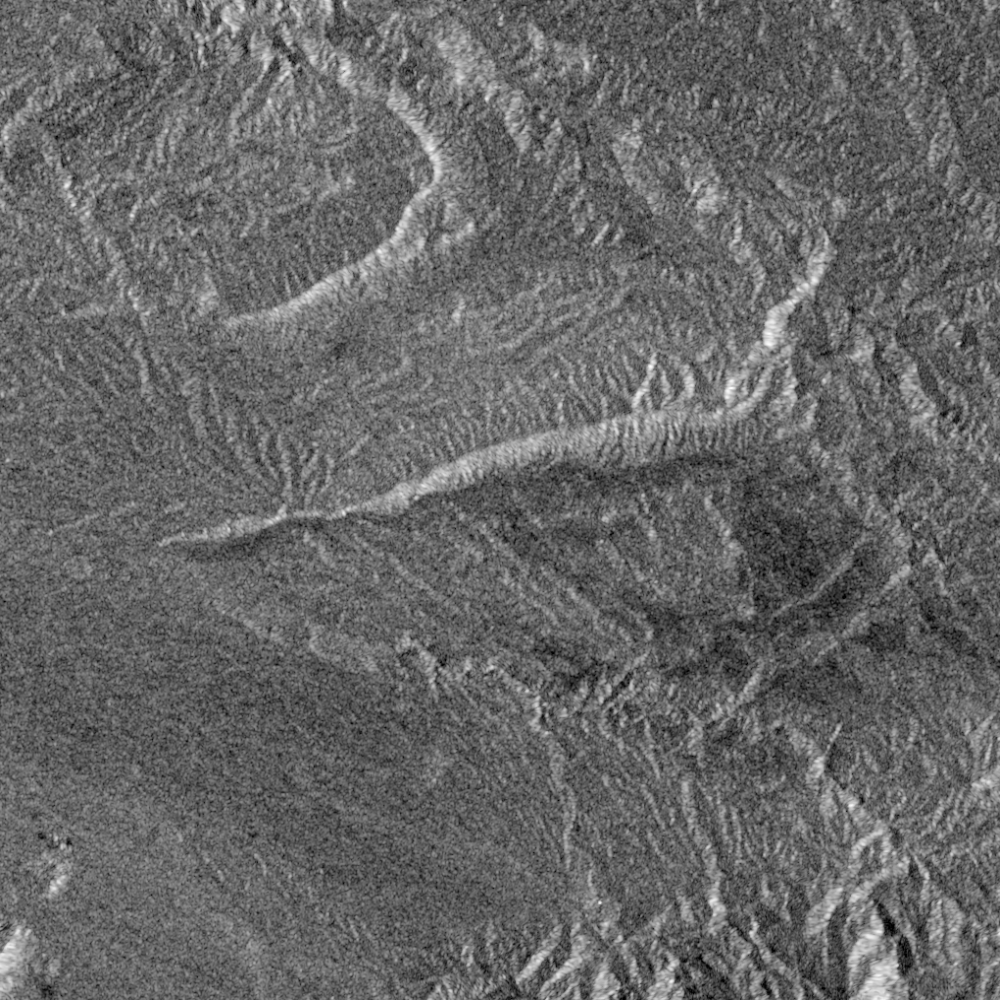}\hspace{0.5cm}
    \includegraphics[scale=.08]{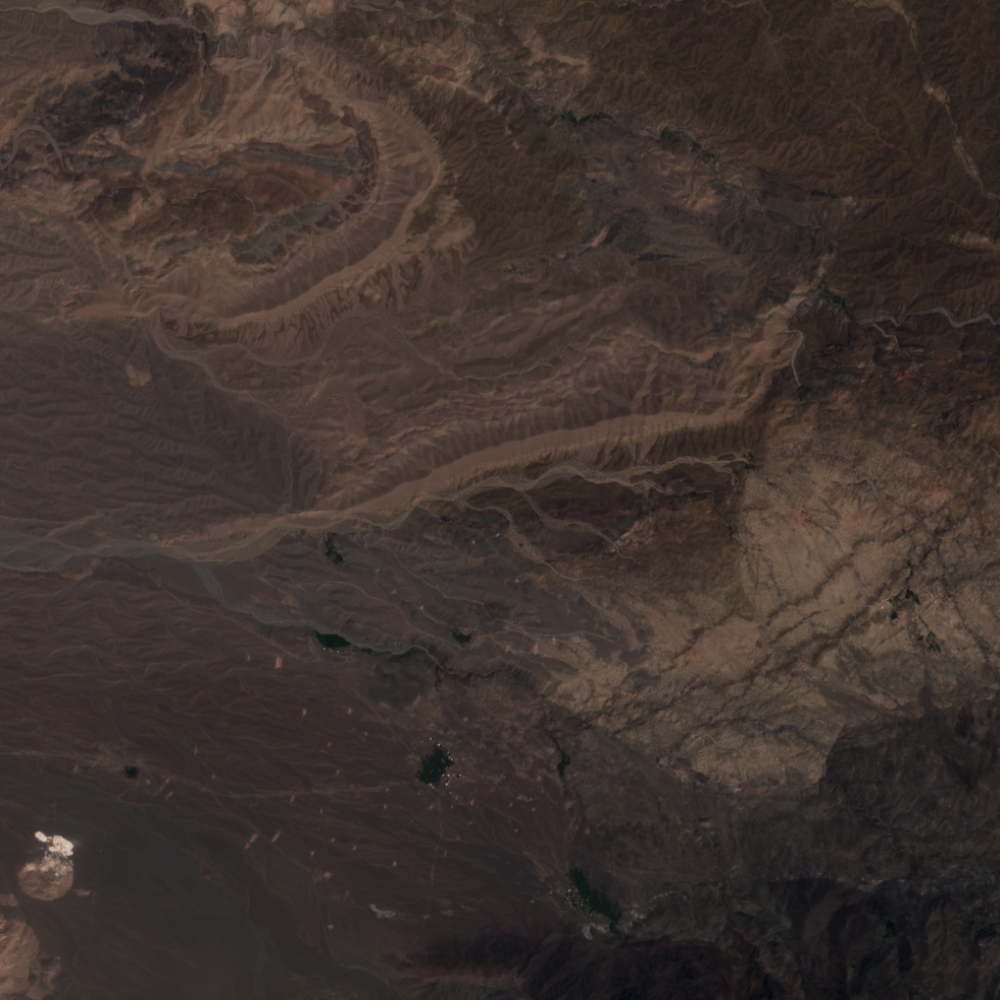}
    \caption{Converter output example. Sentinel-1 on the left and Sentinel-2 on the right.}
    \label{fig:converter_example}
\end{figure}

For Sentinel-1 products the converter first standardizes or normalizes the data to bring them into a range suitable for Machine Learning purposes, then it saves the gray-scale data in png format and with data type uint8 (range 0, 255). 
For Sentinel-2 products the converter normalizes the data, then through the RGB bands it builds a color image and then saves the data in png format with data type uint8 (range 0, 255). An example of converted data is shown in Figure \ref{fig:converter_example}.

Up to now, three of the most common types of normalization techniques are implemented, the min-max, the standardization and the max technique, expressed by equation \ref{eqn:min-max}, \ref{eqn:std} and \ref{eqn:max}, respectively, where "$Image_{in}$" denotes the image to be normalized and "$Image_{out}$" the normalized image. The functions $min$, $max$, $std$ and $mean$ are used to calculate respectively the minimum, the maximum, the standard deviation and the mean of the input image. These are scalar values, built on all the image values. By using the aforementioned equations and the scalars,
the matrix related to the input image is modified  and a new matrix is calculated (the output image)  \cite{patro2015normalization, 4024051}.

\begin{equation}
    Image_{out} = \frac{Image_{in} - min(Image_{in})}{max(Image_{in}) - min(Image_{in})}
    \label{eqn:min-max}
\end{equation}

\begin{equation}
    Image_{out} = \frac{Image_{in} - mean(Image_{in})}{std(Image_{in})}
    \label{eqn:std}
\end{equation}

\begin{equation}
    Image_{out} = \frac{Image_{in}}{max(Image_{in})}
    \label{eqn:max}
\end{equation}

The user can select the previously described conversion mode or set the converter so that it only creates the RGB image and saves both the Sentinel-1 and the Sentinel-2 acquisitions in “.tiff “ format, by avoiding the data normalization in this case.

Normalization in fact is typically used to plot or, in the case of AI applications, to increase effectiveness during learning \cite{data_norm, min-max_norm, features_scaling, DU2002123}, but it is a process that modifies the image value range, and it can be irreversible. An expert user, by selecting the ".tiff" format,  can decide to bypass this step, or even other preprocessing phases, in order to have   raw data to which then applying his customized preprocessing techniques.

\subsection{Cleaner}\label{subs:cleaner}
Unfortunately, it may happen that some images are corrupted or present a too high cloud coverage (in the case of Sentinel-2), therefore the cleaner block has been developed to overcome these problems.

It mainly deals with selecting for each satellite, for each region, for each date, the best image available among the three downloaded every month.

\begin{figure}[!ht]
    \centering
    \includegraphics[scale=.08]{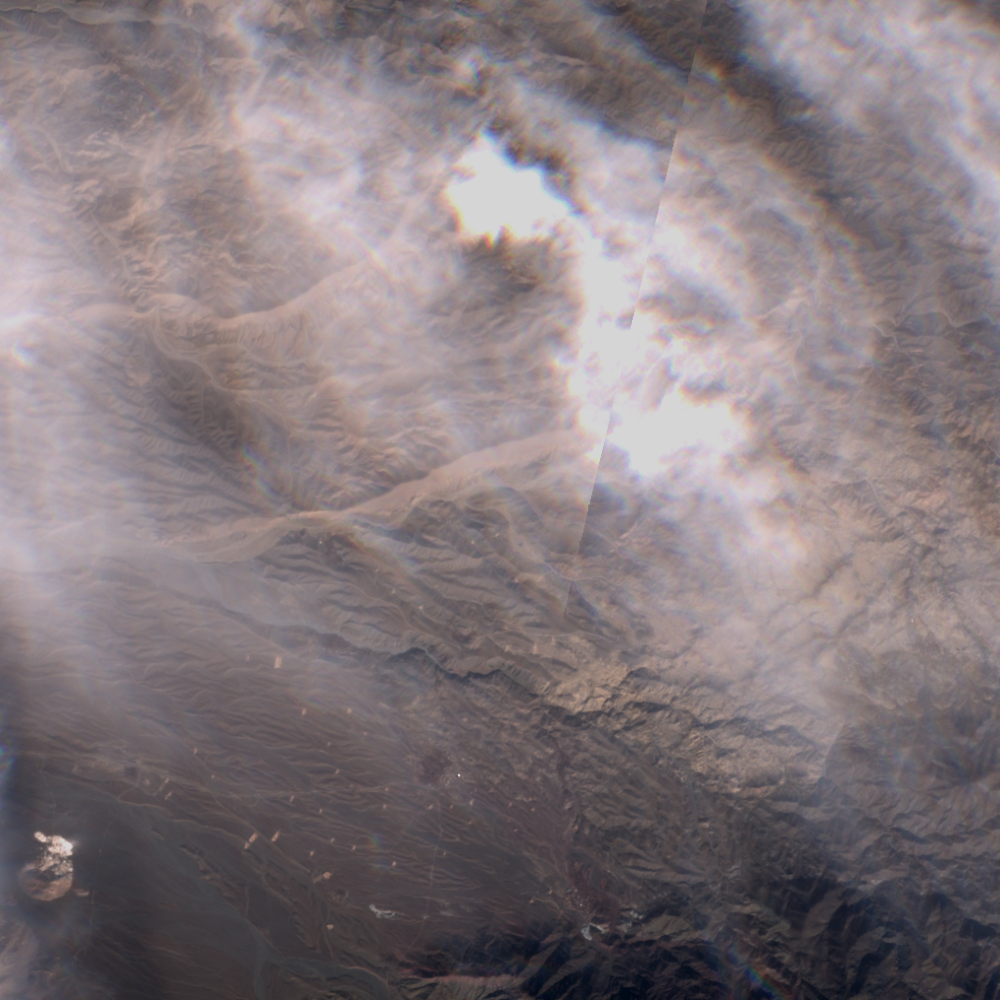}\hspace{0.5cm}
    \includegraphics[scale=.08]{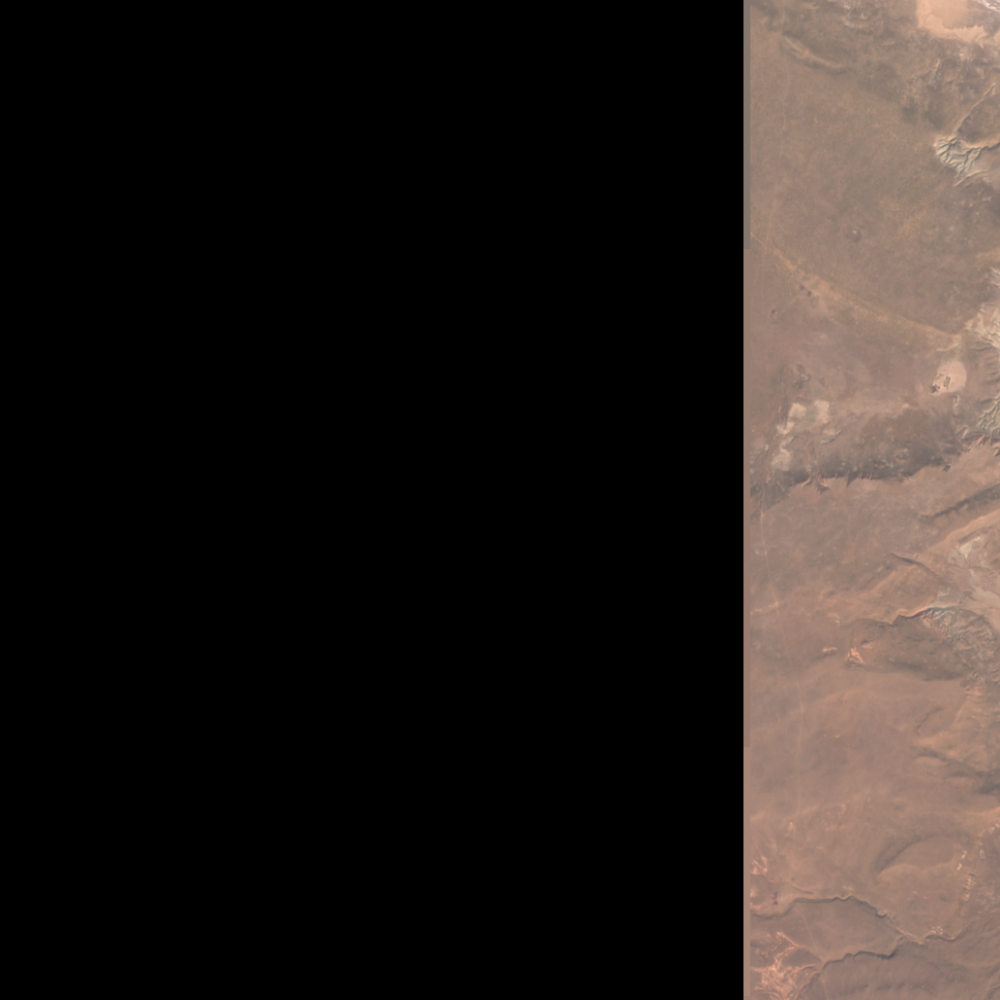}\\~\\
    \includegraphics[scale=.08]{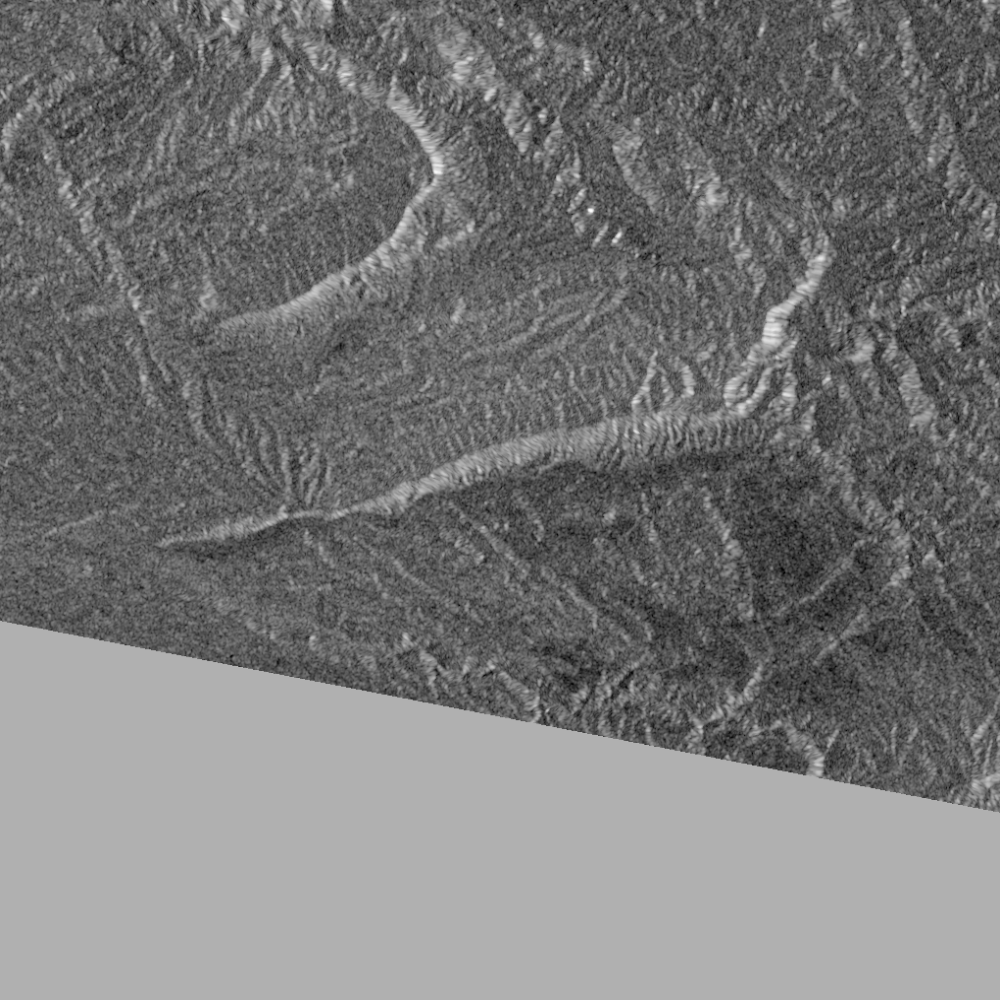}
    \caption{Some "errors" in the downloaded data. Sentinel-2 cloudy on the top left and Sentinel-2 wrong on the top right. Sentinel-1 wrong on the bottom.}
    \label{fig:cleaner_example}
\end{figure}

By using the 3 images (default value), the cleaner selects the best image for each month. In fact there are some "errors" in the downloaded data. For example some Sentinel-1 downloaded data present some black or gray missing parts. Some Sentinel-2 downloaded data present the same problem but in addiction there can be images with a huge cloud coverage. See the images shown in the Figure \ref{fig:cleaner_example} as an example.

For the missing parts or the cloud detection the software uses a threshold. 
It is worth to say that the cleaner is designed, for now, only to remove corrupted or cloudy data, and this is done in an automatic way. Yet, users who want to remove data, with other type of characteristics (for example images acquired over dry areas), can use the cleaner in the manual mode. The manual cleaner allows the user to execute the same functionalities of the automatic cleaner without pre-defined settings. This extra option is available only when   the tool is used in the semi-automatic GUI mode.

\begin{figure}[!ht]
    \centering
    \includegraphics[scale=.34]{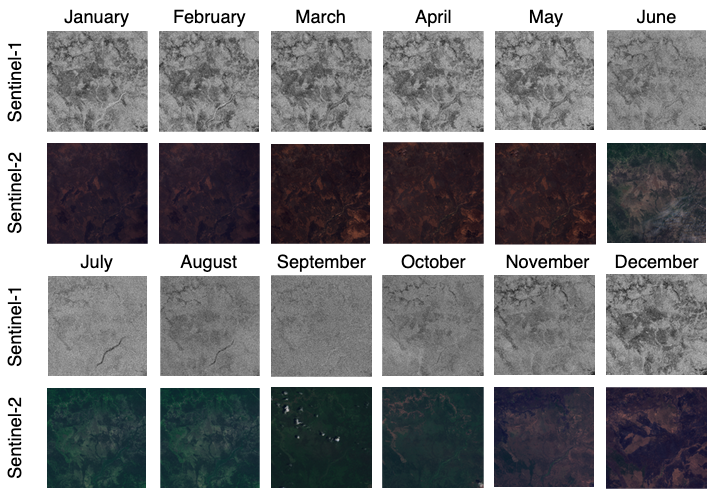}
    \caption{Sample of dataset after the cleaning process}
    \label{fig:dataset_sample}
\end{figure}

At the end of this operation the dataset should be composed, if  the default settings are chosen, of 12 images, one for each month, for each satellite, for a total of 24 images for each region,  free of corrupt or damaged portions, except for some unfavorable cases (see Figure \ref{fig:dataset_sample}). By default the size of each image is 1000x1000 pixels, so from each image it is possible to obtain numerous smaller patches.

\subsection{Patch Extractor}\label{subs:patch_extractor}
The patch extractor is an add-on that extracts smaller images from the final one to increase the samples in the dataset.

During this step, the smaller images are created with the preview of the dataset. Each image contains time series with Sentinel-1 and Sentinel-2 data from a specific geographic region. The images are organized in a matrix form, on the columns there are the different patches extracted, and on the rows there are the different acquisitions over time. 

\begin{figure}[!ht]
    \centering
    \includegraphics[scale=.13]{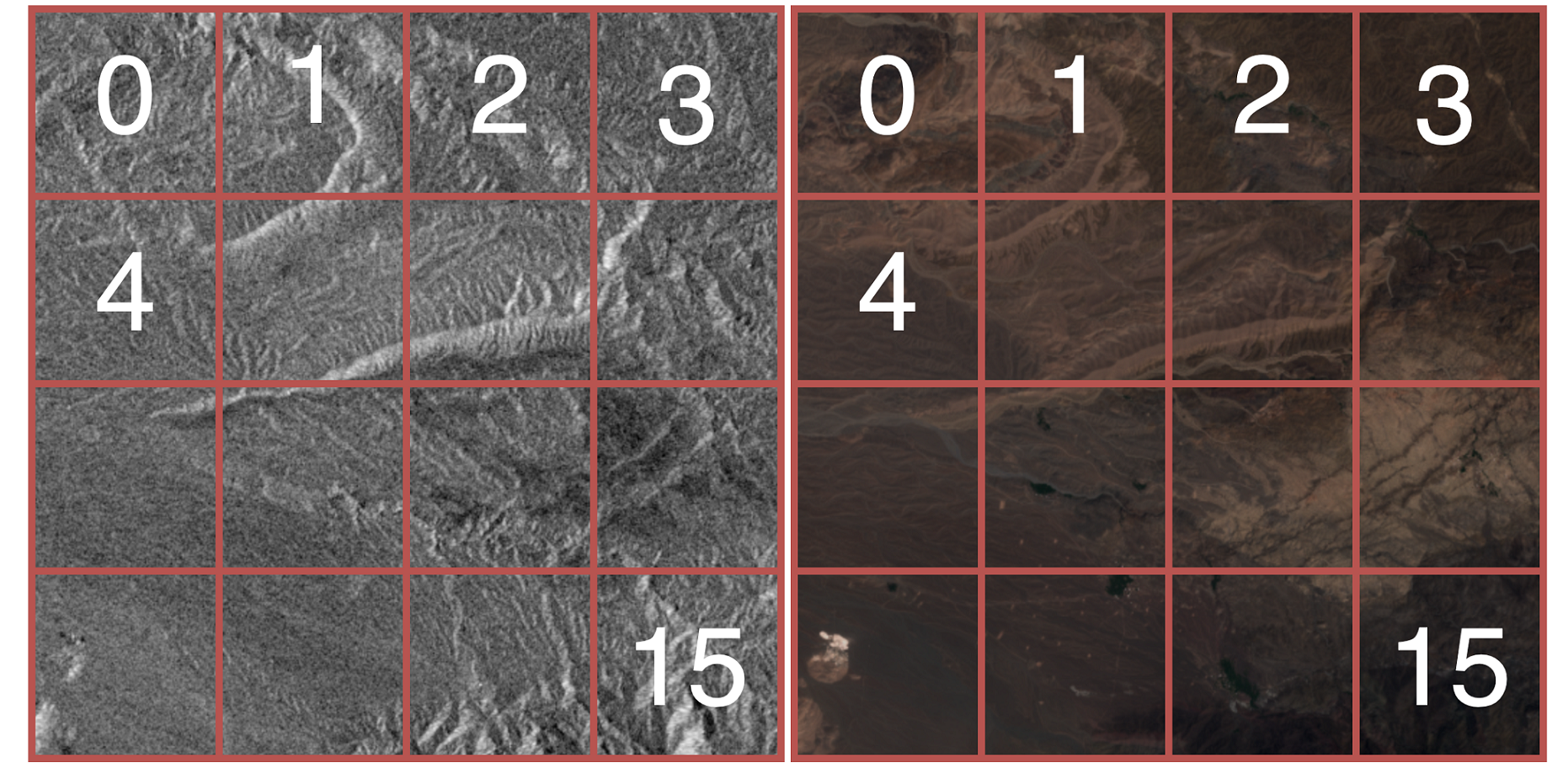}
    \caption{Example of extracted patches from the image after the cleaning step.}
     \label{fig:preview}
\end{figure}

With respect to Figure \ref{fig:dataset_sample}, each Sentinel-1 and Sentinel-2 image after cleaning undergoes the Patch Extractor step, as shown in the Figure \ref{fig:preview}.

The preview can be used to verify all the steps applied by the tool and it can be used also to manually select some data of interest, if necessary. Indeed, the file name contains the information about the position of the images in the dataset. Then using the number of rows and columns, the user can extract a particular portion of the image. 

\section{The Graphical  User  Interface}
\begin{figure}[!ht]
    \centering
    \includegraphics[scale=0.38]{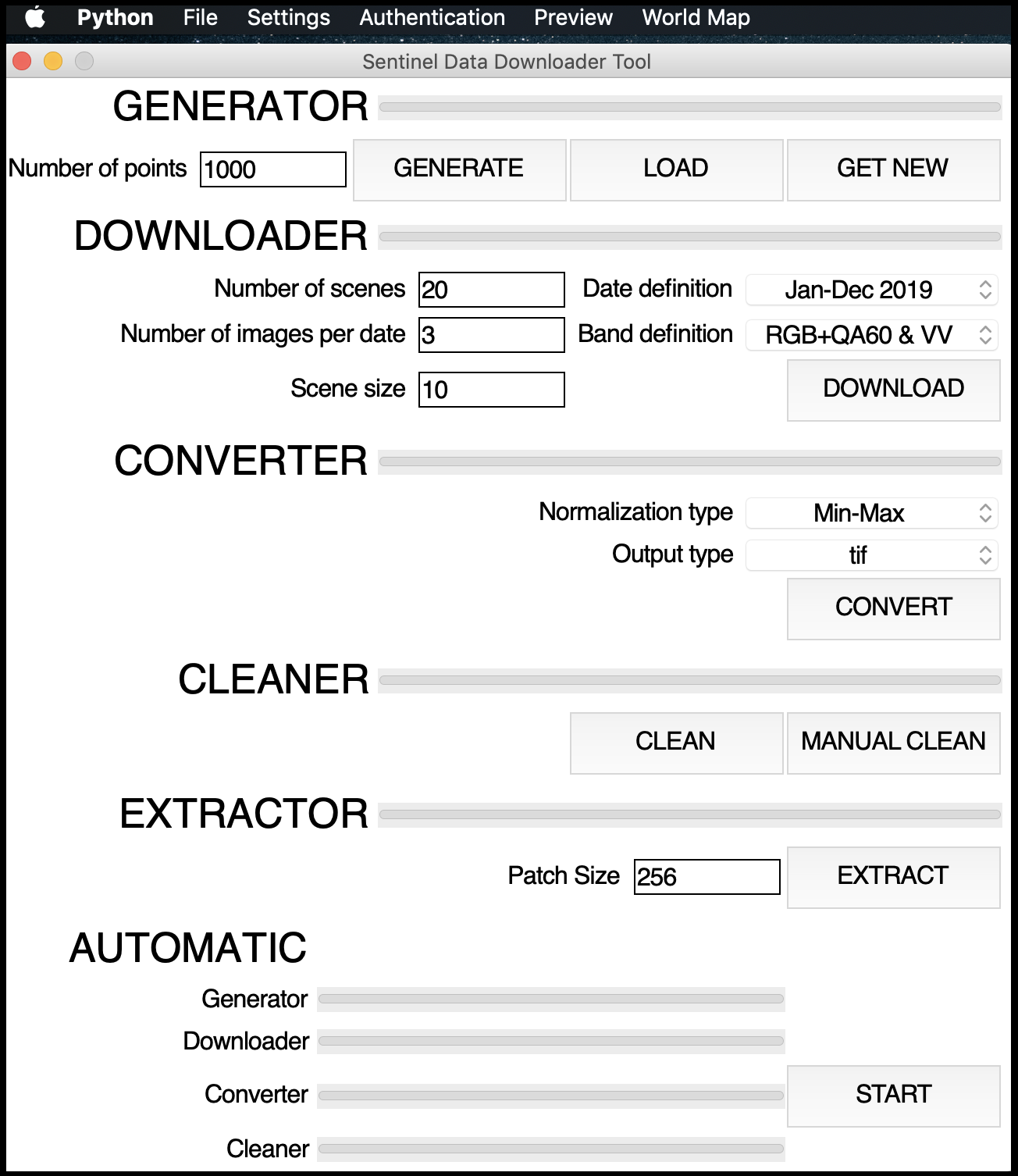}
    \caption{GUI of the Sentinel Data Downloader Tool}
    \label{fig:tool_gui}
\end{figure}

The Graphical  User  Interface (GUI) can be managed by both expert and non-expert users, in fact  two modalities are made available. Expert users can, for instance, utilize  the tool by running the scripts in environment like Jupyter Notebook. Non-expert users, instead, can run the GUI (Graphical User Interface) directly by using the interface shown in the Figure \ref{fig:tool_gui}. Obviously the GUI offers less flexibility with respect to the functioning mode for experts, but it is more intuitive and of easy use.

As it can be seen from the Figure \ref{fig:tool_gui}, the GUI presents two main sections, on the bottom the Automatic section with the START button, and on the remaining part, the possibility of a semi-automatic functioning:

\begin{itemize}
    \item Automatic: the user can run a fully automatic process, using default settings (listed in a setting file), by pressing the START button
    \item Generator, Downloader, Converter, Cleaner, Extractor: the user can change some settings and can run the different processes separately
\end{itemize}

On the top of the Figure \ref{fig:tool_gui}, there are also other options accessible to the user. The most interesting are the Preview and the World Map. Indeed these are two extra components that allow the user to easily navigate thorough the dataset and to plot over a world map the generated points.

The tool and a more detailed user guide can be found on the related Git-Hub page \cite{github_tool}.

\section{Conclusions}
In  this  letter  the  architecture  of  an  innovative tool, enabling researchers to create in an automatic way suitable datasets  for  AI    applications  in  the  EO  context, has been presented.  Two  versions  of  the architecture have been implemented and made available on Git-Hub,  with  a  specific  Graphical  User  Interface  (GUI)  for  non-expert  users. 
For now the tool supports only data available from the Google Earth Engine catalog and it has been fully tested   on Sentinel-1 and Sentinel-2 data. Future work will include the integration of new sources of data and the testing of the tool also in this case. 

\section{Acknowledgments}
The work has been carried out by the University of Sannio   researchers while hosted at the   Phi-Lab of the European Space Research Institute (ESRIN) in Frascati \cite{phi_lab}. A special acknowledgment goes to Pierre-Philippe Mathieu, Chief of ESA ESRIN  Phi-Lab,   for joint brainstorming and sharing of ideas.

\balance
\bibliographystyle{IEEEtran}
\bibliography{IEEEabrv,refs}
~\newline
\end{document}